# Digitally synthesized beat frequency multiplexing for sub-millisecond fluorescence microscopy

Eric D. Diebold[1,3,*], Brandon W. Buckley[1,2,3], Daniel R. Gossett[5], and Bahram Jalali[1,3,4,5]

**Affiliations:**

[1]Department of Electrical Engineering, University of California, Los Angeles, Los Angeles, CA 90095

[2]Department of Physics, University of California, Los Angeles, Los Angeles, CA 90095

[3]California NanoSystems Institute, University of California, Los Angeles, Los Angeles, CA 90095

[4]Department of Surgery, David Geffen School of Medicine, University of California, Los Angeles, Los Angeles, CA 90095

[5]Department of Bioengineering, University of California, Los Angeles, Los Angeles, CA 90095

*ediebold@ucla.edu

Fluorescence imaging is the most widely used method for unveiling the molecular composition of biological specimens. However, the weak optical emission of fluorescent probes and the tradeoff between imaging speed and sensitivity[1] is problematic for acquiring blur-free images of fast phenomena, such as sub-millisecond biochemical dynamics in live cells and tissues[2], and cells flowing at high speed[3]. We report a solution that achieves real-time pixel readout rates one order of magnitude faster than a modern electron multiplier charge coupled device (EMCCD) – the gold standard in high-speed fluorescence imaging technology[4]. Deemed fluorescence imaging using radiofrequency-multiplexed excitation (FIRE), this approach maps the image into the radiofrequency spectrum using the beating of digitally synthesized optical fields. We demonstrate diffraction-limited confocal fluorescence imaging of stationary cells at a frame rate of 4.4 kHz, as well as fluorescence microscopy in flow at a throughput of approximately 50,000 cells per second.

The spatial resolution of modern fluorescence microscopy has been improved to a point such that even sub-diffraction limited resolution is routinely possible [5,6]. However, the demand for continuous, sub-millisecond time resolution using fluorescence microscopy remains largely unsatisfied. Such a real-time fluorescence microscope would enable resolution of dynamic biochemical phenomena such as calcium and metabolic waves in live cells [7,8], action potential sequences in large groups of neurons [9-11], or calcium release correlations and signaling in cardiac tissue [12]. High-speed microscopy is also invaluable for imaging biological cells in flow. Flow imaging can quickly perform high-throughput morphology [3], translocation [13], and cell signaling [14] analysis on large populations of cells. However, as fluorescence imaging frame rates increase towards the kHz range, the small number of photoelectrons generated during each exposure drops below the noise floor of conventional image sensors, such as the charge coupled device (CCD).

The desire to perform high-speed, low-photon number imaging has been the primary driving force behind the development of the EMCCD camera. EMCCDs use on-chip electronic gain to circumvent the high-speed imaging signal-to-noise ratio (SNR) problem. However, while EMCCDs can exhibit 1000-fold gain, the serial pixel readout strategy limits the full frame (512x512 pixels) rate to less than 100 Hz. A photomultiplier tube (PMT) can provide 1000x higher gain, 10x lower dark noise, and 50x higher bandwidth than EMCCDs, but they are not typically manufactured in large array format. This limits the utility of PMTs in fluorescence microscopy to point-scanning applications [15]. PMT-based single point laser scanning fluorescence microscopes are capable of high sensitivity at pixel readout rates similar to EMCCDs, but the serial beam scanning ultimately limits its speed. Recently, scientific complementary metal-oxide-semiconductor (sCMOS) cameras have become popular for many fluorescence imaging applications, but their lack of an internal gain mechanism still renders the EMCCD the preeminent solution for the lowest light and highest frame rate fluorescence imaging applications [16].

Here, we present a high-speed radiofrequency (RF) communications approach to fluorescence microscopy, which combines the benefits of PMT sensitivity and speed with frequency-domain signal multiplexing, RF spectrum digital synthesis, and digital lock-in amplification to enable fluorescence imaging at kHz frame rates. FIRE microscopy is a frequency-domain analog of serial time-encoded amplified microscopy (STEAM) [17]. STEAM maps pixels into optical wavelengths, and multiplexes the pixels into a serial time-domain stream. STEAM also uses optical image amplification and time stretching to enable real-time, MHz frame rate brightfield imaging. However, its wavelength encoding of pixels renders STEAM incompatible with fluorescence imaging. In contrast, frequency-domain excitation approaches to fluorescence imaging have previously been employed for lifetime spectroscopy [18,19].

The central feature of FIRE is its ability to excite fluorescence in each individual point of the sample at a distinct radiofrequency. Digitally synthesized radiofrequency "tagging" of pixels' fluorescence emission occurs at the beat frequency between two interfering, frequency-shifted laser beams. Similar to frequency-domain multiplexing in wireless communication systems, each pixel of a FIRE image is assigned its own RF frequency. A single-element PMT simultaneously detects fluorescence from multiple pixels, and an image is reconstructed from the frequency components of the PMT output. In a two-dimensional FIRE image, pixels are analogous to points on a time-frequency Gabor lattice, as illustrated in Figure 1.

FIRE performs beat frequency excitation multiplexing by employing acousto-optic devices in a Mach-Zehnder interferometer (MZI) configuration. As shown in Figure 1, the light in one arm of the MZI is frequency shifted by a 100-MHz bandwidth acousto-optic deflector (AOD), driven by a comb of RF frequencies, phase-engineered [20] to minimize its peak-to-average power ratio. The AOD produces multiple deflected optical beams possessing a range of both output angles and frequency shifts [21]. Light in the second arm of the interferometer passes through an acousto-optic frequency shifter, driven by a single RF tone, which provides a local oscillator (LO) beam. A cylindrical lens is used to match the LO beam's angular divergence to that of the RF comb beams. At the MZI output, the two beams are combined and focused to a horizontal line on the sample, mapping frequency shift to space. Since fluorescent molecules in the sample function as square-law detectors of the total optical field, fluorescence is excited at the various beats defined by the difference frequencies of the two arms of the interferometer. Fluorescence emission from the sample is detected by a PMT in a descanned confocal configuration, using a slit aperture to reject out-of-plane fluorescence emission. A resonant scan mirror performs high-speed scanning in the transverse direction for two-dimensional imaging. Given the finite frequency response of fluorophores, the LO beam frequency shift is chosen to heterodyne the beat frequency excitation spectrum to baseband, in order to maximize the useable modulation bandwidth. This is necessary because AODs typically operate over an upshifted, sub-octave passband to avoid harmonic interference [21]. Direct digital synthesis (DDS) of the RF comb used to drive the AOD defines each pixel's excitation by a specific radiofrequency and phase, resulting in phase coherence between the RF comb and the detected signal [18]. This phase coherence enables image de-multiplexing using a parallel array of phase-sensitive lock-in amplifiers, implemented in Matlab. FIRE's parallel readout results in a maximum pixel rate equal to the bandwidth of the AOD.

Figure 2 illustrates the operational principle of FIRE microscopy. Immobilized 15-µm diameter fluorescent polystyrene beads are imaged at a frame rate of 4.4 kHz, using 256 excitation frequencies spaced by 300 kHz, for a total pixel readout rate of 76.8 MHz. The detected time-domain signal (Fig. 2A) offers no indication of the lateral bead location, yet a short-time Fourier transform (STFT) of three windows of the signal (Fig. 2B) indicate the different frequency components associated with the positions of each bead. The vertical locations of the beads are recovered from the reference output of the 2.2-kHz resonant scan mirror, and the final image is formed (Fig. 2C).

To demonstrate FIRE microscopy on biological samples, we imaged adherent cells stained with various fluorophores at a frame rate of 4.4 kHz. NIH 3T3 mouse embryonic fibroblasts, C6 astrocyte rat glial fibroblasts, and *S. cerevisiae* yeast were stained with a fluorescent cytosol stain (Calcein AM) or nucleic acid stain (Syto16). Figure 3 shows image crops of cells, taken with both the FIRE microscope and a conventional widefield fluorescence microscope based on a 1280x1024 pixel CMOS camera. The frame rate difference of nearly 3 orders of magnitude is mitigated by the electronic gain of the PMT detector combined with digital lock-in amplifier image de-modulation. Measuring the single pixel sensitivity of FIRE, we found the limit of detection in a 100-kHz bandwidth to be $51 \times 10^{-12}$ M of Rhodamine B (Supplementary Figure S1). To further demonstrate the ability of two-dimensional FIRE to record fluorescent dynamic phenomena, we imaged fluorescent beads flowing at a velocity of 2.3 mm/s inside a microfluidic channel (Supplementary Video 1).

Flow cytometry is another application of high-speed fluorescence measurement. As compared to single-point flow cytometry, imaging flow cytometry provides a myriad of information that can be particularly useful for high-throughput rare cell detection [22]. The high flow velocities associated with flow cytometry demand fast imaging shutter speeds and high sensitivity photodetection to generate high SNR, blur-free images. Commercial imaging flow cytometers available from Amnis Corporation use time delay and integration CCD technology in order circumvent this issue, but the serial pixel readout strategy of this technology currently limits the device to a throughput of approximately 5,000 cells per second [23,24]. However, to overcome the Poisson statistics inherent to detecting rare circulating tumor cells in milliliter samples of blood, higher throughput is needed [25,26].

To demonstrate high-speed imaging flow cytometry using FIRE, we used a single stationary linescan of 125 pixels spaced by 800 kHz (pixel readout rate of 100 MHz), and imaged Syto16-stained MCF-7 human breast carcinoma cells, flowing in a microfluidic channel at a velocity of 1 m/s. Assuming a cell diameter of 20 $\mu$m, this velocity corresponds to a maximum throughput of 50,000 cells per second. For comparison, we also imaged Syto16-stained MCF-7 cells in flow at the same velocity using a frame transfer EMCCD in single exposure mode (512x512 pixels). As shown in Figure 4, while the high sensitivity and gain of the EMCCD yields a reasonable SNR image at the camera's minimum exposure time of 10 $\mu$s, image blur still remains at these flow velocities. In contrast, the FIRE linescan shutter speed of 1.25 $\mu$s yields blur-free images with comparable SNR.

Other approaches to frequency-domain fluorescence imaging have been reported previously [19,27]. However, the kHz-bandwidth, mechanical modulation schemes used in these works do not provide sufficient pixel readout rates required for sub-millisecond imaging. This implementation of FIRE features pixel readout rates in the 100 MHz range, but this rate could be directly extended to more than 1 GHz through the use of wider bandwidth AODs [21]. FIRE's maximum modulation frequency, and thus maximum pixel readout rate, is intrinsically limited by the sample's fluorescence lifetime. If the excitation frequency is less than $1/\tau_l$, where $\tau_l$ is the fluorescence lifetime of the sample, the emitted fluorescence will oscillate at the excitation frequency with an appreciable modulation [1]. Further, beat-frequency modulation is critical to the speed of the technique; the beating of two coherent, frequency-shifted optical waves produces a single RF tone, without any harmonics that can introduce pixel crosstalk and reduce the useable bandwidth.

The flexibility afforded by digitally synthesizing the RF spectrum provides complete, real-time control over the number of pixels, pixel frequency spacing, and field of view. Additionally, FIRE's use of DDS to engineer the amplitude and phase of each individual pixel excitation allows both equalization of non-uniform AOD diffraction efficiency, and improvement of the image dynamic range. Specifically, the ability to phase-engineer the excitation frequency comb enables the dynamic range of each pixel to scale as $D/\sqrt{N}$, where $D$ is the dynamic range of the PMT, and $N$ is the pixel-multiplexing factor [28]. This is in contrast to the case where all excitation frequencies' initial phases are locked, which yields images with a dynamic range of $D/N$. While FIRE fundamentally presents a tradeoff in dynamic range for speed, it improves in sensitivity when compared to single point scanning fluorescence microscopy. For equivalent frame rates, multiplexing the sample excitation by a factor of $N$ yields an $N$-fold increase in the dwell time of each pixel. This consequently improves FIRE's shot noise-limited image SNR by a factor of $\sqrt{N}$.

Beat frequency multiplexing is also applicable to other types of laser scanning microscopy, including two-photon excited fluorescence microscopy. Perhaps most notably, FIRE is inherently immune to pixel crosstalk arising from fluorescence emission scattering in the sample – the effect that typically limits the imaging depth in multifocal multiphoton microscopy [29]. In combination with fast fluorophores [30], FIRE microscopy may ultimately become the technique of choice for observation of nano- to microsecond-timescale phenomena using fluorescence microscopy.

References:

1   Lakowicz, J. R. *Principles of fluorescence spectroscopy*. 3rd edn, (Springer, 2006).
2   Scanziani, M. & Hausser, M. Electrophysiology in the age of light. *Nature* **461**, 930-939, doi:Doi 10.1038/Nature08540 (2009).
3   Elliott, G. S. Moving Pictures: Imaging Flow Cytometry for Drug Development. *Comb Chem High T Scr* **12**, 849-859 (2009).
4   Coates, C. New sCMOS vs. Current Microscopy Cameras. *Andor White Paper* (2011).
5   Hell, S. W. & Wichmann, J. Breaking the Diffraction Resolution Limit by Stimulated-Emission - Stimulated-Emission-Depletion Fluorescence Microscopy. *Opt Lett* **19**, 780-782 (1994).
6   Rust, M. J., Bates, M. & Zhuang, X. W. Sub-diffraction-limit imaging by stochastic optical reconstruction microscopy (STORM). *Nat Methods* **3**, 793-795, doi:Doi 10.1038/Nmeth929 (2006).
7   Tsien, R. Y. Fluorescent Indicators of Ion Concentrations. *Method Cell Biol* **30**, 127-156 (1989).
8   Petty, H. R. Spatiotemporal chemical dynamics in living cells: From information trafficking to cell physiology. *Biosystems* **83**, 217-224, doi:Doi 10.1016/J.Biosystems.2005.05.018 (2006).
9   Grinvald, A., Anglister, L., Freeman, J. A., Hildesheim, R. & Manker, A. Real-Time Optical Imaging of Naturally Evoked Electrical-Activity in Intact Frog Brain. *Nature* **308**, 848-850 (1984).
10  Cheng, A., Goncalves, J. T., Golshani, P., Arisaka, K. & Portera-Cailliau, C. Simultaneous two-photon calcium imaging at different depths with spatiotemporal multiplexing. *Nat Methods* **8**, 139-U158, doi:Doi 10.1038/Nmeth.1552 (2011).
11  Yuste, R. & Denk, W. Dendritic Spines as Basic Functional Units of Neuronal Integration. *Nature* **375**, 682-684 (1995).
12  Shiferaw, Y., Aistrup, G. L. & Wasserstrom, J. A. Intracellular Ca-2 waves, afterdepolarizations, and triggered arrhythmias. *Cardiovasc Res* **95**, 265-268, doi:Doi 10.1093/Cvr/Cvs155 (2012).
13  George, T. C. *et al.* Quantitative measurement of nuclear translocation events using similarity analysis of multispectral cellular images obtained in flow. *J Immunol Methods* **311**, 117-129, doi:Doi 10.1016/J.Jim.2006.01.018 (2006).
14  Khalil, A. M., Cambier, J. C. & Shlomchik, M. J. B Cell Receptor Signal Transduction in the GC Is Short-Circuited by High Phosphatase Activity. *Science* **336**, 1178-1181, doi:Doi 10.1126/Science.1213368 (2012).
15  Pawley, J. B. in *Handbook of Biological Confocal Microscopy*   (Springer, New York, 2006).
16  Coates, C. New sCMOS vs. Current Microscopy Cameras. *Andor white paper* (2011).
17  Goda, K., Tsia, K. K. & Jalali, B. Serial time-encoded amplified imaging for real-time observation of fast dynamic phenomena. *Nature* **458**, 1145-U1180, doi:Doi 10.1038/Nature07980 (2009).
18  Lakowicz, J. R. & Berndt, K. W. Lifetime-Selective Fluorescence Imaging Using an Rf Phase-Sensitive Camera. *Rev Sci Instrum* **62**, 1727-1734 (1991).
19  Howard, S. S., Straub, A., Horton, N. G., Kobat, D. & Xu, C. Frequency-multiplexed in vivo multiphoton phosphorescence lifetime microscopy. *Nat Photon* **7**, 33-37, doi:http://www.nature.com/nphoton/journal/v7/n1/abs/nphoton.2012.307.html - supplementary-information (2013).
20  Golay, M. J. E. Complementary Series. *Ire T Inform Theor* **7**, 82-&, doi:Doi 10.1109/Tit.1961.1057620 (1961).

21	Young, E. H. & Yao, S. K. Design Considerations for Acoustooptic Devices. *P Ieee* **69**, 54-64 (1981).
22	Goda, K. *et al.* High-throughput single-microparticle imaging flow analyzer. *P Natl Acad Sci USA* **109**, 11630-11635, doi:Doi 10.1073/Pnas.1204718109 (2012).
23	https://http://www.amnis.com/imagestream.html, 2012.
24	Basiji, D. A., Ortyn, W.E. Imaging and analyzing parameters of small moving objects such as cells. United States patent 6,211,955 (2001).
25	Allard, W. J. *et al.* Tumor cells circulate in the peripheral blood of all major carcinomas but not in healthy subjects or patients with nonmalignant diseases. *Clin Cancer Res* **10**, 6897-6904, doi:Doi 10.1158/1078-0432.Ccr-04-0378 (2004).
26	Nagrath, S. *et al.* Isolation of rare circulating tumour cells in cancer patients by microchip technology. *Nature* **450**, 1235-U1210, doi:Doi 10.1038/Nature06385 (2007).
27	Wu, F. *et al.* Frequency division multiplexed multichannel high-speed fluorescence confocal microscope. *Biophys J* **91**, 2290-2296, doi:Doi 10.1529/Biophysj.106.083337 (2006).
28	Nee, R. v. & Prasad, R. *OFDM for wireless multimedia communications.* (Artech House, 2000).
29	Kim, K. H. *et al.* Multifocal multiphoton microscopy based on multianode photomultiplier tubes. *Opt Express* **15**, 11658-11678, doi:Doi 10.1364/Oe.15.011658 (2007).
30	Kralj, J. M., Douglass, A. D., Hochbaum, D. R., Maclaurin, D. & Cohen, A. E. Optical recording of action potentials in mammalian neurons using a microbial rhodopsin. *Nat Methods* **9**, 90-U130, doi:Doi 10.1038/Nmeth.1782 (2012).

**Acknowledgements**

We thank Prof. Dino Di Carlo (UCLA) for use of his lab's cell culture facilities. The authors would like to acknowledge the Broad Stem Cell Research Center at UCLA 2012 Innovation Award for financial support. We thank Dr. Laurent Bentolila for assistance with the EMCCD imaging which was performed at the California NanoSystems Institute Advanced Light Microscopy/Spectroscopy Shared Facility at UCLA.

**Author contributions**

E.D.D. conceived of the beat frequency multiplexing approach, built the FIRE microscope, and collected the data. B.W.B. conceived of and implemented the data processing algorithms, engineered the DDS excitation frequency combs and performed image processing. D.R.G. cultured and stained the biological samples, and fabricated microfluidic channels. B.J. conceived of the use of DDS and other communication techniques for FIRE, and supervised the project. E.D.D. wrote the first draft of the manuscript, and all authors contributed to subsequent revisions.

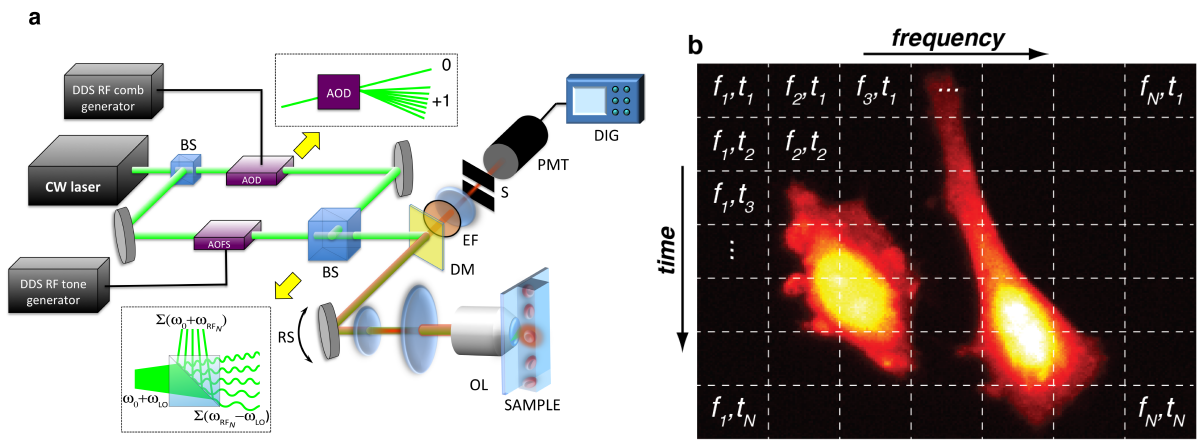

**Figure 1. (A)** Schematic diagram of the FIRE microscope. BS: beamsplitter AOD: acousto-optic deflector AOFS: acousto-optic frequency shifter DM: dichroic mirror EF: fluorescence emission filter OL: objective lens PMT: photomultiplier tube DIG: 250 MS/s digital recording oscilloscope RS: resonant scanning mirror. Upper inset: the AOD produces a single diffracted 1$^{st}$ order beam for each RF comb frequency. Lower inset: beat frequency generation at the MZI output. Not shown: A cylindrical lens is placed after the AOFS to match the divergence of the LO beam to that of the RF beams. **(B)** Gabor lattice diagram of FIRE's frequency-domain multiplexing approach. Points in the horizontal direction are excited in parallel at distinct radiofrequencies. Scanning this linescan excitation in the vertical direction using a galvanometer generates a two-dimensional image.

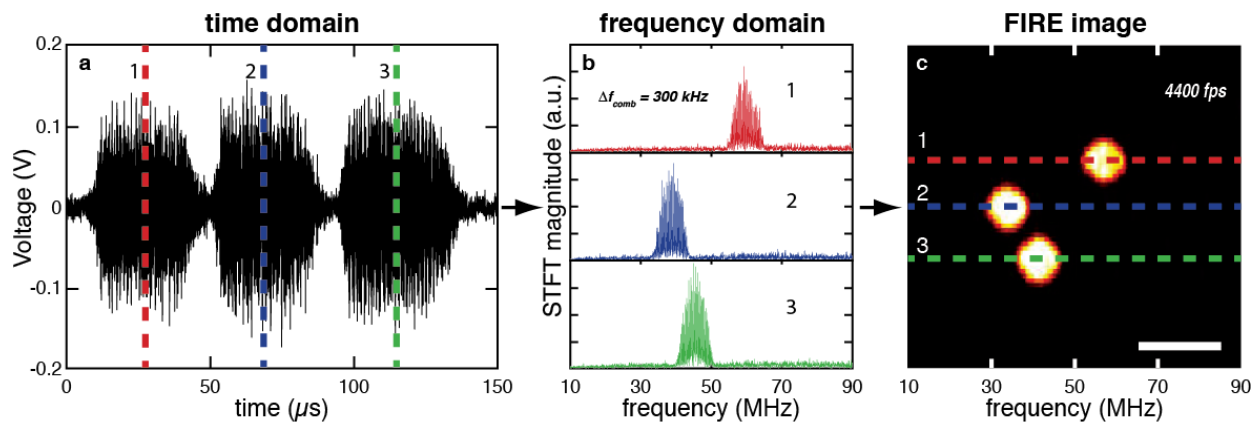

**Figure 2.** Illustration of the FIRE concept. **(A)** Time domain data output from the PMT. **(B)** Short-time Fourier transforms of the signal in (A), indicating the bead horizontal positions. **(C)** 256x256 pixel image of three immobilized fluorescent beads recorded using 256 excitation frequencies. The sample was imaged at a 4.4-kHz frame rate. The vertical axis in the image is oversampled to yield 256 pixels. Scale bar = 30 μm.

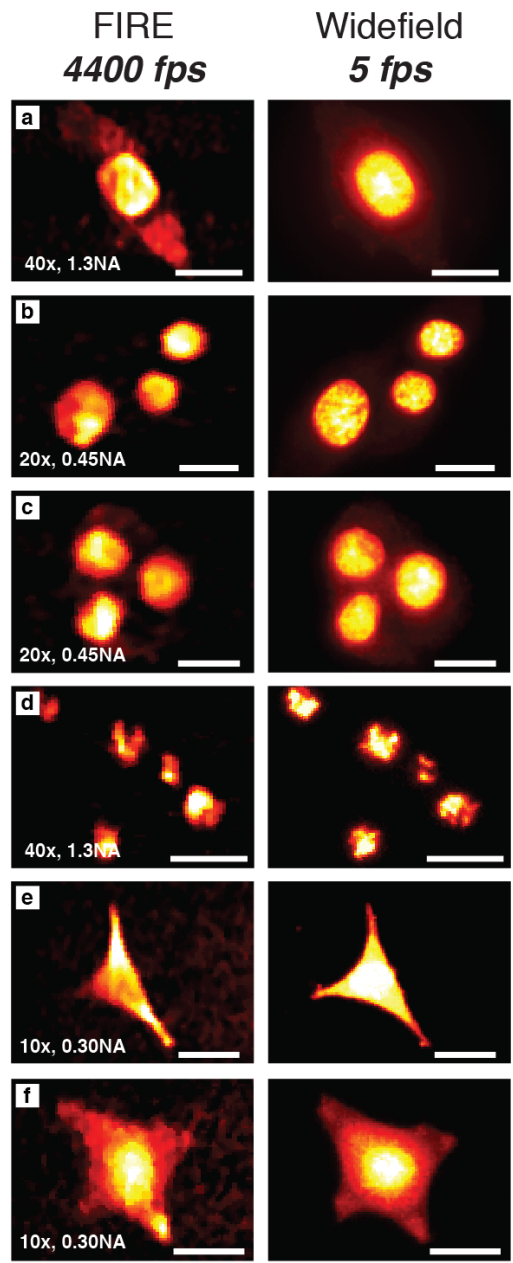

**Figure 3.** Comparison of FIRE microscopy with widefield fluorescence imaging. 488nm laser excitation was used for FIRE imaging (8.5μW per pixel, measured before the objective), and mercury lamp excitation was used for widefield imaging. All FIRE images use an RF comb frequency spacing of 400 kHz, and are composed of 200x92 pixels. Slight vignetting is observed in the FIRE images due to the mismatch of the Gaussian profile of the LO beam with the RF comb beams. This mismatch and the resulting vignetting can be eliminated using digital pre-equalization of the RF comb in the DDS generator. The particular objective lens used is denoted in each FIRE image. **(A-C)** C6 astrocytes stained with Syto16. Scale bars = 10 μm. **(D)** *S. cerevisiae* yeast stained with Calcein AM. Scale bars = 5 μm. **(E,F)** NIH 3T3 cells stained with Calcein AM. Scale bars = 20 μm.

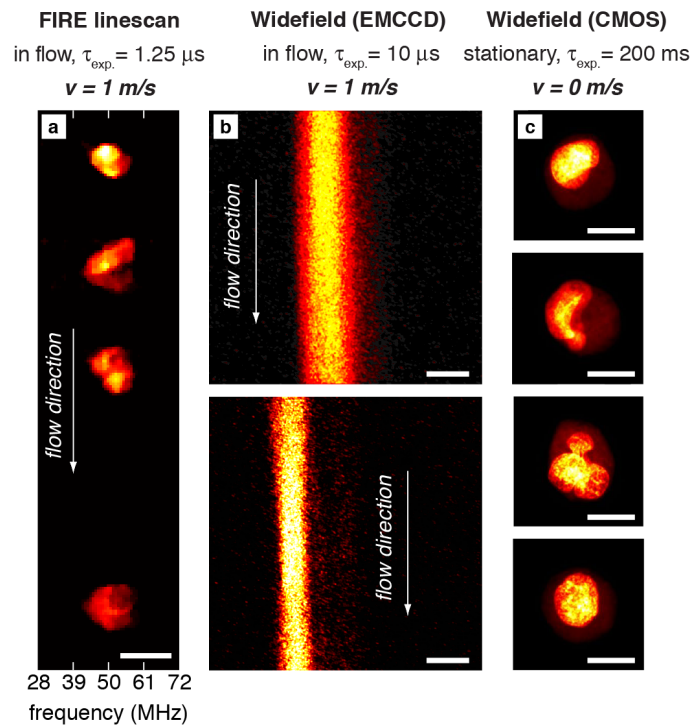

**Figure 4.** High-speed imaging flow cytometry. All images are of MCF-7 breast carcinoma cells, stained with Syto16, taken using a 60x, 0.70-NA objective lens. **(A)** Representative FIRE images of cells flowing at a velocity of 1 m/s, taken using a pixel frequency spacing of 800 kHz, and 54 μW of 488-nm laser power per pixel, measured before the objective. **(B)** Single 10-μs exposure frame transfer EMCCD images of individual cells flowing at a velocity of 1 m/s. The electron multiplier gain was adjusted to produce maximal image SNR, and the EMCCD vertical shift time was set to 564 ns. **(C)** Representative widefield fluorescence images of stationary MCF-7 cells. All scale bars are 10 µm.

## Supplementary Information:

**Photodetection and digitization of fluorescence signals:**

Fluorescence emission is detected by a bialkali PMT (R3896, Hamamatsu) after passing through a dichroic mirror and a bandpass filter. The current signal from the PMT is amplified by a 400-MHz bandwidth current amplifier with 5 kV/A transimpedance gain. This voltage signal is digitized using a 250 MS/s, 8-bit resolution oscilloscope. For 2-D scanned images, the reference output from the resonant scan mirror is used for triggering, and is also digitized and saved for image reconstruction. For the flow cytometry experiments, the digitizer is triggered off of the image signal itself. The digitized data is then transferred to a PC for signal processing.

**Digital Signal Processing:**

Direct digital synthesis of the LO and RF comb beams is accomplished using the two outputs of a 5 GS/s arbitrary waveform generator, which are amplified to drive the AOD and AOFS. After photodetection and digitization, two digital de-multiplexing algorithms can be used to recover the fluorescence image from the frequency-multiplexed data. The first employs the short time Fourier transform (STFT) and is similar to the demodulation technique used in orthogonal frequency division multiplexing (OFDM) communication systems. The second technique uses an array of digital lock-in amplifiers to heterodyne and de-multiplex each comb-line separately.

The STFT method works by segmenting the data sequence into frames, and performing a discrete Fourier transform (DFT) on each frame to recover the frequency-multiplexed signal. Each frame corresponds to a 1D line-scan. Following standard procedures of OFDM demodulation, in order to avoid pixel-pixel cross talk from power spreading across frequency bins, the time duration of each frame is set as an integer multiple of the inverse of the frequency comb spacing. In this case, the frequency channels are said to be orthogonal, and the DFT bins lie precisely on the frequency comb lines. The maximum line rate in this scenario is equal to the frequency comb spacing.

The lock-in amplifier demodulation technique is implemented by digitally mixing the data signal with copies of each frequency comb line. Mixing of the comb lines and the signal downshifts the corresponding comb line to baseband. A low pass filter (LPF) extinguishes all other comb lines, leaving only the modulated fluorescent signal from the comb line of interest. To obviate phase locking the reference to the signal, both in-phase (I) and quadrature phase (Q) mixing terms are calculated. The magnitude of the sum of I- and Q-channels are equal to the amplitude of the signal. The bandwidth of the LPF is less than half the comb spacing to prevent pixel crosstalk in frequency space. With the reduced analog bandwidth after filtering, each pixel's signal can be boxcar-averaged and under-sampled to at least the Nyquist rate, equal to the frequency comb spacing. The under-sampled data rate corresponds to the line rate of the system.

Although both the DFT and lock-in technique can be used to de-multiplex the fluorescence image, the lock-in technique has certain advantages: (a) there is no orthogonality requirement, leaving more flexibility in comb line configuration, (b) the nominal line rate is determined by the under-sampling factor, allowing for line-rates above the minimum Nyquist rate, and (c) the reference and signal can be phase locked, either by *a priori* estimation of the signal phase or via deduction from the ratio of I and Q channels. Phase locked operation rejects out of phase noise, resulting in a 3-dB improvement in SNR.

**Two-dimensional image scanning**

In this implementation of FIRE, images are acquired on an inverted microscope, in a descanned configuration, using a 2.2 kHz resonant scan mirror. To avoid nonlinear space-to-time mapping from the resonant scanner in the vertical direction, an aperture is placed in the intermediate image plane of the imaging system to limit the sample excitation to the approximately linear portion of the scan field. A sine correction algorithm is further applied to the image in Matlab to compensate for any residual distortion in the image arising from the sinusoidal deflection pattern of the mirror. Processing of the raw computed images is performed in Matlab. Brightness and contrast adjustments, thresholding, and 2-dimensional low-pass filtering are performed on all images.

**Imaging Flow Cytometry**

MCF-7 breast carcinoma cells were stained with Syto16, prior to fixation with formaldehyde. The cells were then suspended in phosphate-buffered saline, and flowed through a linear, rectangular cross-section 110μmx60μm microfluidic channel made from polydimethylsiloxane, using a syringe pump, at a fixed volumetric rate. The fluid flow velocity is calculated using the equation

$$V = \frac{Q}{A}$$

where Q is the volumetric flow rate, and A is the cross-sectional area of the channel. Vertical scaling calibration of the images is performed after imaging 10μm spherical beads flowing at the same volumetric flow rate.

**FIRE design criteria - spatial resolution, number of pixels, and field of view:**

The FIRE microscope is a diffraction-limited technique. Like other laser scanning microscopy techniques, the minimum transverse spatial resolution is the diffraction limited spot size determined by the numerical aperture of the objective and the laser excitation wavelength.

The pixel rate of the system is equal to the product of the nominal line rate and the number of pixels per line. With the nominal line rate limited to the comb spacing, the maximum pixel rate of FIRE is equal to the bandwidth of the AOD,

$$B_{AOD} = (p_x \times p_y) \times r_{frame},$$

where $p_x$ and $p_y$ are the number of pixels in the x and y dimensions, and $r_{frame}$ is the image frame rate.

The number of resolvable points per frequency-multiplexed line is determined by the time-bandwidth product of the AOD. To satisfy the Nyquist sampling criterion, the number of pixels per line should be at least twice this value,

$$p_x = 2 \times TBP_{AOD},$$

where we have chosen the x-axis as the AOD deflection direction.

The field of view in each direction is the number of resolvable points times the diffraction limited spot-size, d,

$$FOV_x = d \times TBP_{AOD}$$

$$FOV_y = d \times \frac{p_y}{2},$$

in accordance with the Nyquist sampling criterion. The range of the scan mirror deflection in the y-direction should be chosen to match the $FOV_y$ above.

**Sensitivity measurement:**

To quantify the sensitivity of the FIRE system, we measured the limit of detection of a single pixel. Dilutions of Rhodamine B in deionized water ranging from 10pM to 100nM were used as test samples. 10μL drops of each sample were placed inside cells consisting of a microscope slide, a 120μm-thick spacer, and a number 1.5 coverslip. A sample of pure deionized water was used as the reference sample. 1 mW of 532-nm laser power excited the sample through a 40x, 1.3-NA, oil-immersion microscope objective. We chose the beat frequency of the pixel to be 40 MHz, which lies in the middle of the AOD bandwidth. The PMT was set to a voltage of -1000 V for all measurements. The PMT output signal was amplified by a 50-dB gain low noise amplifier, and digitized by an 8-bit, 250-MS/s oscilloscope. Phase-sensitive lock-in detection was applied to the signal in Matlab to reject out-of-phase noise in the system. The data points in Figure S1 each represent the mean and standard deviations of 100,000 sample values taken at a particular concentration. Using a linear fit to the data, the limit of detection (3σ above the reference background mean) was determined to be $51 \times 10^{-12}$ M. The use of high-frequency lock-in detection effectively circumvents 1/f noise present in the system.

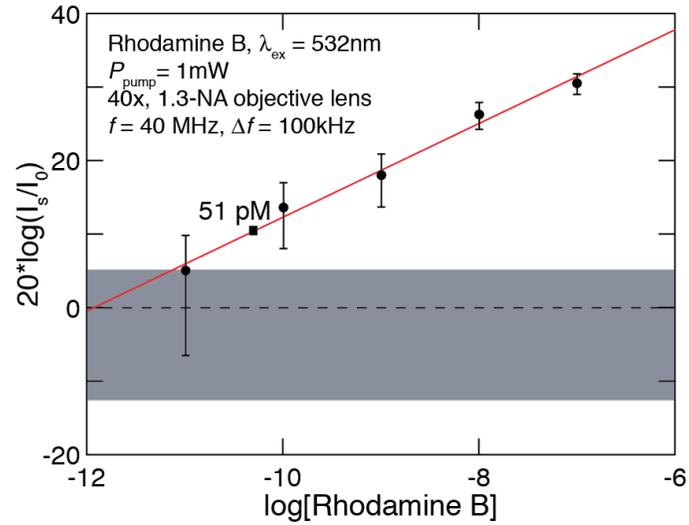

Figure S1. Single pixel sensitivity of the FIRE microscope using phase-sensitive lock-in detection. The limit of detection, calculated from the mean background noise level recorded from a sample of deionized water, is $51 \times 10^{-12}$ M in a 100 kHz bandwidth. The dashed line and gray shaded area represent the mean and standard deviation of the signal taken from the reference sample.

**Cell culture:**

NIH3T3 and MCF-7 cells and C6 astrocytes were propagated in Dulbecco's Modified Eagle Medium with 10% fetal bovine serum and 1% penicillin streptomycin at 37°C and 5% $CO_2$. Liquid cultures of *Saccharomyces cerevisiae* were grown in tryptic soy broth at 240 rpm and 37°C.

**Cell staining:**

Prior to staining, cultured mammalian cells were released from culture flasks, seeded on glass slides, and allowed to spread for 24 hours. Mammalian cells were stained with either 4 µM Syto16 green fluorescent nucleic acid stain in phosphate buffered saline (PBS) for 30 minutes, 1 µM Calcein Red-Orange AM in culture media for 45 minutes, or 1 µM Calcein AM in culture media for 45 minutes. Cells were washed twice with PBS then fixed for 10 minutes with 4% paraformaldehyde in PBS. Following fixation, cells were washed twice with PBS and prepared for either stationary or flow-through microfluidic imaging. For stationary imaging, number 1.5 cover glasses were placed on slides and sealed. In an effort to preserve the shape of adhered mammalian cells for flow-through microfluidic experiments a cell scraper was used to remove spread, stained, and fixed cells from glass slides. The cells were diluted in PBS.

*S. cerevisiae* were stained in suspension using the same concentration of Calcein AM for 45 minutes, washed twice with PBS, fixed with 4% paraformaldehyde in PBS, and washed twice with PBS. For stationary imaging the cells were seeded on glass slides and sealed under number 1.5 coverslips.

**Microfluidic channel fabrication:**

Microfluidic channels were fabricated using standard photolithographic and replica molding methods. Briefly, a mold was fabricated out of KMPR 1025 (MicroChem Corp., Newton, MA) on a silicon wafer using photolithography. Silicone elastomer (Sylgard 184, Dow Corning, Corp., Midland, MI) was cast and cured on the mold according to technical specifications. The cured elastomer was peeled off the mold. Vias for tubing interfaces were bored in the elastomer replica. A glass slide and the molded side of the elastomer replica were activated with air plasma, brought into contact, and baked at 65°C for 3 hours to bond the materials and form the microchannels. Tubing was inserted into the vias. For flow imaging, cell suspensions were introduced into the channel at a fixed volumetric rate using a syringe pump.